\documentclass[aps,pra,twocolumn,longbibliography,superscriptaddress,amsmath,amssymb,amsfonts,citeautoscript,floatfix,10pt]{revtex4-2}
\usepackage{graphicx}
\usepackage{bm}
\usepackage{color}
\usepackage{amsmath}
\usepackage{amssymb}
\usepackage[urlcolor=blue,colorlinks=true,citecolor=blue,linkcolor=blue,pdfstartview={FitH},bookmarks=false]{hyperref}
\usepackage{soul}
\usepackage{cancel}
\usepackage{float}
\usepackage{tabularx}
\graphicspath{{figures/}{./figures/}{.}}

\newcommand{\blambda}{{\bm\lambda}}
\newcommand{\bsigma}{{\bm\sigma}}
\newcommand{\btau}{{\bm\tau}}
\newcommand{\mH}{\mathcal{H}}
\newcommand{\T}{\mathcal{T}}
\newcommand{\C}{\mathcal{C}}
\newcommand{\mP}{\mathcal{P}}
\newcommand{\V}{\mathcal{V}}
\newcommand{\im}{\textrm{i}}
\newcommand{\e}{\textrm{e}}
\setstcolor{red}

\begin{document}

\title{Unconventional topological transitions in a self-organized magnetic ladder}

\author{Maciej M. Ma\'ska}
\email[e-mail:]{maciej.maska@pwr.edu.pl}
\affiliation{Department of Theoretical Physics, Wroc{\l}aw University of Science and Technology, 50-370 Wroc{\l}aw, Poland}

\author{Nicholas Sedlmayr}
\email[e-mail:]{sedlmayr@umcs.pl}
\affiliation{Institute of Physics, M. Curie-Sk\l{}odowska University, 20-031 Lublin, Poland}

\author{Aksel Kobia\l{}ka}
\affiliation{Institute of Physics, M. Curie-Sk\l{}odowska University, 20-031 Lublin, Poland}

\author{Tadeusz Doma\'nski}
\affiliation{Institute of Physics, M. Curie-Sk\l{}odowska University, 20-031 Lublin, Poland}

\date{\today}

\begin{abstract}
It is commonly assumed that topological phase transitions in topological superconductors are accompanied by a closing of the topological gap or a change of the symmetry of the system. We demonstrate that an unconventional topological phase transition with neither gap closing nor a change of symmetry is possible. We consider a nanoscopic length ladder of atoms on a superconducting substrate, comprising self-organized magnetic moments coupled to itinerant electrons. For a range of conditions, the ground state of such a system prefers helical magnetic textures, self-sustaining topologically nontrivial phase. Abrupt changes in the magnetic order as a function of induced superconducting pairing or chemical potential can cause topological phase transitions without closing the topological gap. Furthermore, the ground state prefers either parallel or anti-parallel configurations along the rungs, and the anti-parallel configuration causes an emergent time reversal asymmetry protecting Kramer's pair's of Majorana zero modes, but in a BDI topological superconductor. We determine the topological invariant and inspect the boundary Majorana zero modes.
\end{abstract}

\maketitle


\section{Introduction}

In recent years, the investigation of topological phases~\cite{Haldane-2017} has deepened our understanding of many-body systems and predicted emergence of novel states of matter~\cite{HasanKane-2010,QiZhang-2011}. Prior to the discovery of topological phases the conventional Landau picture classified phase transitions into discontinuous or continuous ones, associated with symmetry breaking and emergence of order. This understanding has been deepened by the investigation of quantum phase transitions~\cite{Sachdev-2000}, and analogies can been drawn with dynamical quantum phase transitions~\cite{Heyl-2018}. Berezinski-Kosterlitz-Thouless transitions~\cite{Kosterlitz-2017} are one example which lies beyond the Landau paradigm and describe the condensation of {\it topological} defects.

In contrast to continuous transitions, which can be understood from the behavior of order parameters associated with symmetry breaking, topological phase transitions originate from a change of the topology of the ground state, which is protected by a gap. Ground states can be classified topologically by which can be deformed into each other by symmetry preserving unitary deformations, and they can be characterized by topological invariants. Such topological band insulators and superconductors are hence further classified according to their symmetries \cite{Ryu-2010,Kruthoff-2017}. Furthermore the bulk topological invariants are related to the number of boundary modes emerging inside a topological gap~\cite{Teo2010}.

\begin{figure}[t!]
\includegraphics[width=0.47\textwidth]{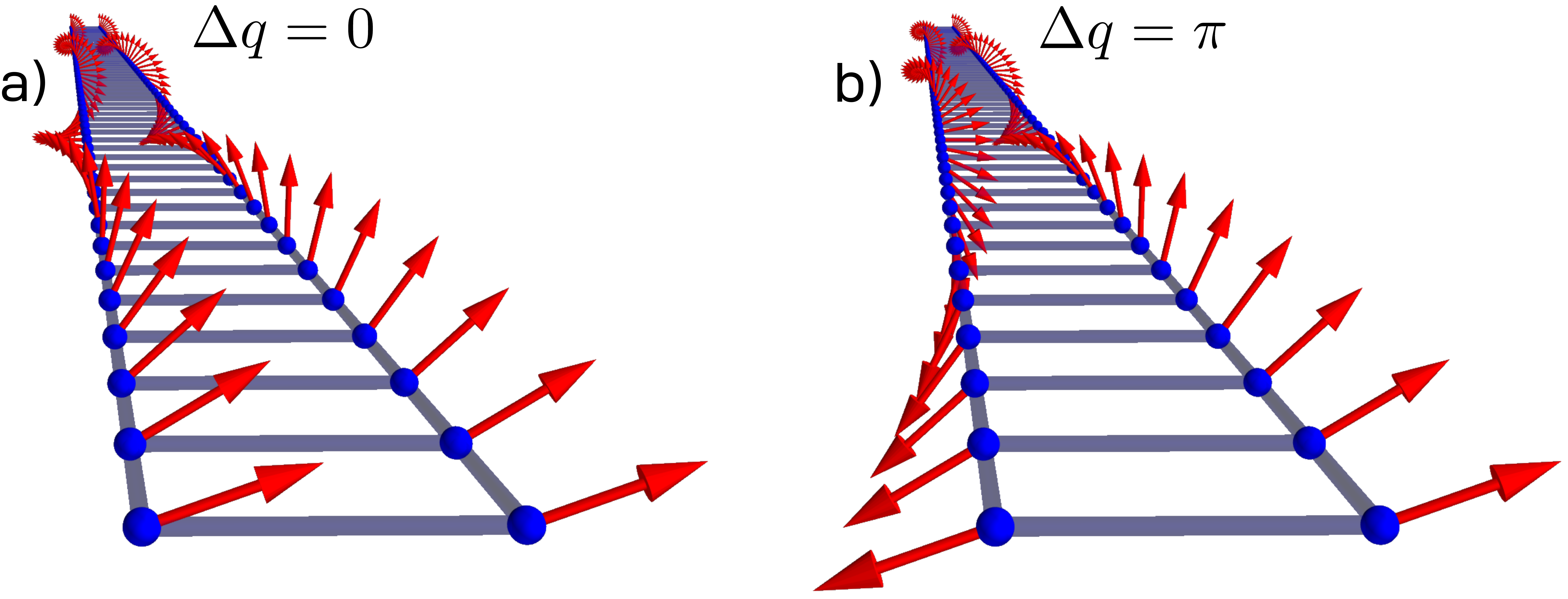}
\caption{Schematic view of the magnetic ladder whose spiral order has parallel ($\Delta q=0$) or anti-parallel ($\Delta q= \pi$) alignment in the rungs, which are decisive for the topologically non-trivial superconducting phase.}
\label{scheme}
\end{figure}

Topological phase transitions therefore require either a closing of the gap or a change of the symmetry \cite{Ezawa2013,Yang_2013}.
Here, we propose an exotic type of topological transition accompanied neither by a change 
of the symmetry nor by a closing of the topological gap. We discuss a specific realization of such
an unconventional topological phase transition in a nanoscopic magnetic ladder proximitized 
to bulk superconductor. The mechanism by which the unconventional topological phase transition occurs is generic, and could be generalized to any dimension or topological symmetry class. This is in contrast to the first order topological phase transition introduced in Ref.~\cite{Juricic2017} which exists only in 3D topological insulators and is driven by a discontinuous change in the magnetization of the ground state as a function of applied magnetic field. In our case it is rather the magnetic ordering of the ground state which has discontinuities in the parameter space. The discontinuity in the magnetic order as a function of a parameter leads to a topological phase transition without the topological gap closing or the symmetry changing. In other contexts magnetic ordering has already been shown to be important for topological superconductivity, from non-collinear magnetic order in chains~\cite{Ojanen-2014,Kim2014a}, magnet-superconductor hybrid structures \cite{Crawford2020} to the Majorana zero modes (MZMs) confined at skyrmions~\cite{Garnier2019,Diaz2021}.

\begin{figure*}
\begin{center}
\includegraphics[width=\textwidth]{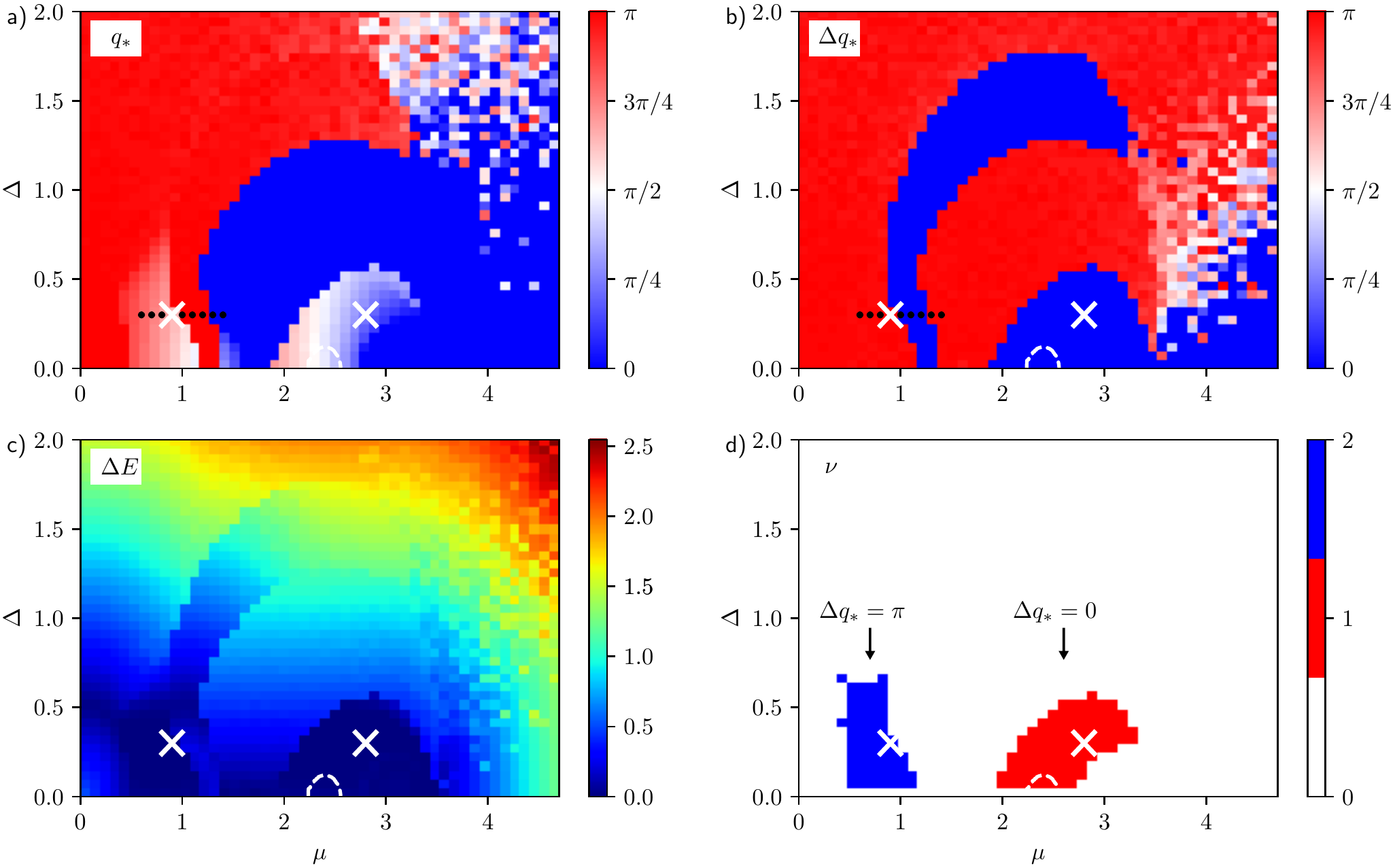}
\end{center}
\caption{The upper left (a) and right (b) panels show $q_*$ and $\Delta q_*$ with respect to $\Delta$ and $\mu$.  The lower left panel (c) presents the quasiparticle energy gap, $\Delta E$, around the Fermi level. In particular $\Delta E=0$ (dark regions) {\it may} imply existence of the zero-energy modes of topologically nontrivial phase. The lower right panel (d) shows a ${\mathbb Z}$ topological invariant. The white crosses indicate the points 
for which we present the quasiparticle spectra in Fig.~\ref{fig:spectr1}. Energy landscapes at points marked by black dots in (a) and (b) panels (for $\Delta=0.3$, $\mu=0.6\to 1.4)$ are presented in Fig.~\ref{fig:landscape}.}
\label{fig:phase}
\end{figure*}

The model we consider has a further interesting property as it is an example of a chiral BDI topological superconductor which nonetheless has an additional time reversal symmetry in some regions of the phase diagram, in which case it should be more correctly thought of as AIII. Thus we find Kramer's pairs of MZMs even though the $\mathbb{Z}$ topological invariant of the BDI class remains valid, which becomes equal to 2 to reflect the existence of the multiple edge modes~\cite{Sedlmayr-2015JPCM}. The MZMs themselves~\cite{Kitaev2001,Oreg2010,Lutchyn2011a,Oreg2010} are of interest both due to their fundamental properties, such as their non-Abelian braiding, and the possible application of this property for fault tolerant quantum computation~\cite{Kitaev2003,Beenakker2020}.


\section{Microscopic model}

Topological superconductivity in magnetic ladders driven by strong Rashba coupling and a Zeeman field has been previously considered by several groups~\cite{Ojanen-2014,Klinovaja-2014,Paaske-2014,Loss-2014,Schrade-2017,Reeg-2017,Hsu-2018,Dmytruk-2019,Haim-2019,Schulz-2019,Klinovaja-2020}. Here, we investigate a different scenario due to the self-organization of the classical moments deposited on a superconducting substrate, and its ability to automatically develop topological phases. Magnetic order on a substrate can re-order as parameters are varied~\cite{Braunecker-2015,Paaske-2016,Scalettar-2017,PhysRevLett.102.116403,PhysRevB.80.165119,PhysRevB.87.235427}, and the self-sustained topological superconductivity of single magnetic chains~\cite{Vazifeh-2013,Simon-2013,Braunecker-2015,Paaske-2016,Scalettar-2017} has been predicted to survive to experimentally accessible temperatures  \cite{Klinovaja-2013,Maska-2019}. We show that the magnetic moments of the ladder prefer to align either ferro- or antiferro-magnetically, without any topological superconductivity, or they develop two types of chiral arrangements both of which imply the topologically non-trivial phase (Fig.~\ref{scheme}). Surprisingly, the transition to these topological phases is not necessarily related to a closing and reopening of the protecting gap.

Electrons moving on $2\times N$ sites of the magnetic ladder can be described by the tight-binding Hamiltonian
\begin{align}
    H=-&t\sum_{i,\sigma}\left(\sum_j\hat{c}^\dagger_{i,j,\sigma}\hat{c}_{i+1,j,\sigma}
    +\hat{c}^\dagger_{i,1,\sigma}\hat{c}_{i,2,\sigma}
    + {\rm H.c.}\right)\nonumber\\
    -&\mu\sum_{i,j,\sigma}\hat{c}^\dagger_{i,j,\sigma}\hat{c}_{i,j,\sigma}
    +J\sum_{i,j}{\bm S}_{i,j}\cdot\hat{\bm s}_{i,j}\nonumber \\
    +&\Delta\sum_{i,j} \left(\hat{c}^{\dagger}_{i,j,\uparrow}\hat{c}^{\dagger}_{i,j,\downarrow}
    +{\rm H.c.}\right),
    \label{eq:hamil}
\end{align}
where $i=1,2,\ldots N$ enumerates the sites along the legs, and $j\in\{1,2\}$ refers to the legs. We use the conventional notation for the annihilation (creation) operators $\hat{c}_{i,j,\sigma}$ ($\hat{c}_{i,j,\sigma}^{\dagger}$)
and define the spin operator
\begin{equation}
\hat{\bm s}_{i,j}=\frac{1}{2}\sum_{\alpha,\beta}\hat{c}^\dagger_{i,j,\alpha}{\bm \sigma}_{\alpha\beta}
\hat{c}_{i,j,\beta}
\end{equation}
with ${\bm\sigma}$ being a vector of the Pauli matrices.
We assume that electrons interact with the magnetic moments ${\bm S}_i$ whose slow dynamics can be treated classically. Such local moments can be expressed in spherical coordinates as
\begin{equation}
    {\bm S}_{i,j} = S\left(\sin\theta_{i,j}\cos\phi_{i,j},\:\sin\theta_{i,j}\sin\phi_{i,j},\:\cos\theta_{i,j}\right)
\end{equation}
in terms of the polar and azimuthal angles $\theta_i$ and $\phi_i$, respectively.
Here we focus on the coplanar spin configuration $\theta_i=\pi/2$, i.e.~assuming 
${\bm S}_{i,j} = S\left(\cos\phi_{i,j},\:\sin\phi_{i,j},\:0 \right)$. We take $t$ as the energy unit throughout the paper. Note, that despite the classical approximation for the localized spins the spins and the electrons are coupled and this coupling determines the properties of the entire system by inducing the ordering in the magnetic subsystem and driving the electrons to a topological state.

We have found that at zero temperature, the magnetic moments eventually develop a perfect spiral ordering
$\phi_{i,1}  = iq$, $\phi_{i,2} = iq + \Delta q$,
where $q$ is the spiral pitch and $\Delta q$ is the phase difference between the legs. Note that $q$ describes the ordering along the legs (i.e., $q=0$ corresponds to ferromagnetic and $q=\pi$ to antiferromagnetic order), whereas $\Delta q$ describes the relative phase between the spirals on the two legs. In other words, $\Delta q$ amounts for the phase difference along the rungs. In the following, we self-consistently compute the values $q_{*}$ and $\Delta q_{*}$  of $q$ and $\Delta q$, which minimize the ground state energy. 

\begin{figure}[t]
\begin{center}
\includegraphics[width=0.45\textwidth]{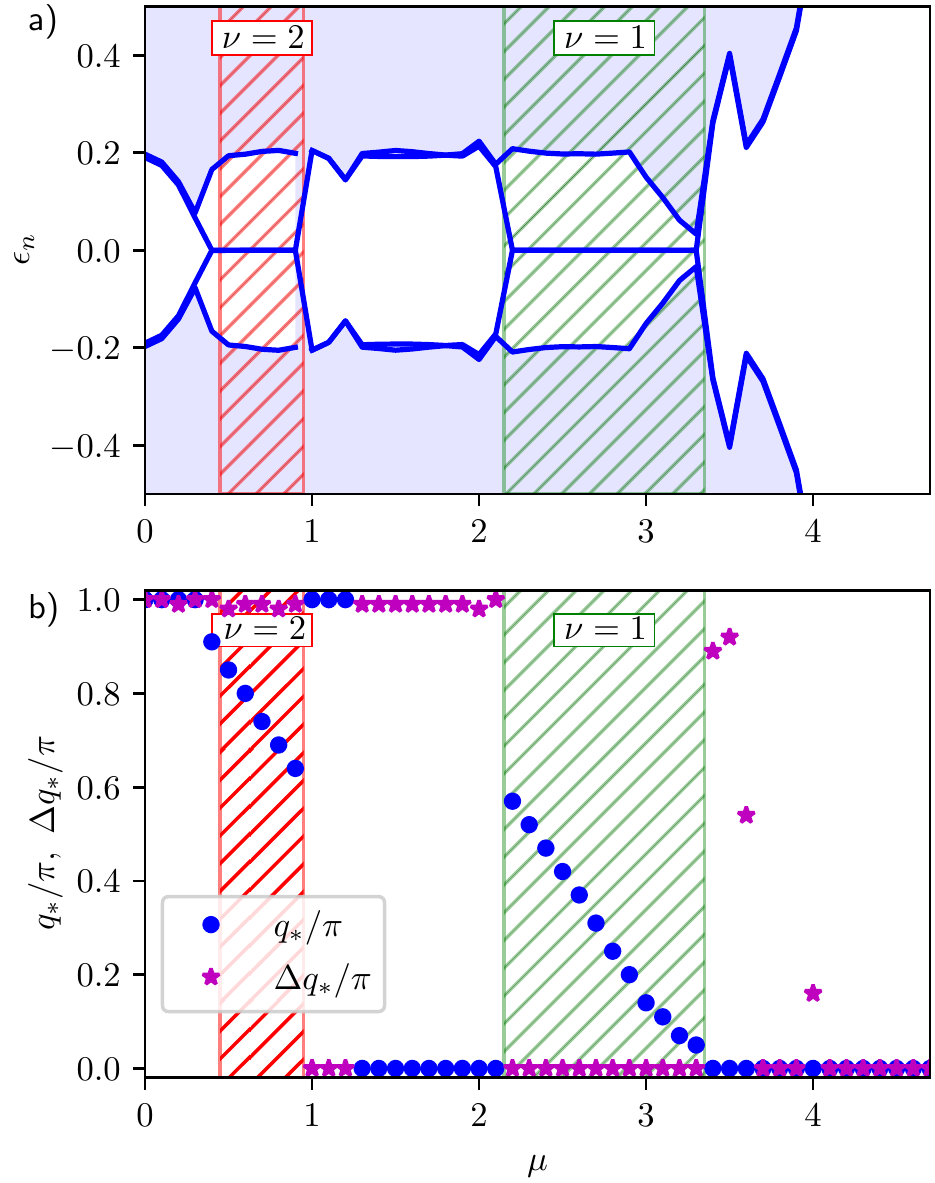}
\end{center}
\caption{Eigenenergies $\epsilon_{n}$ and the angles $q_{*}$, $\Delta q_{*}$ obtained for a line through the phase diagrams with $\Delta=0.3$. The green and red hatched area show topological regions with $\nu=1$ and $\nu=2$, respectively. Note that only the phase transition near $\mu\approx 3.3$ is accompanied by a gap closing. At all other transitions one of the angles suddenly shifts, indicating an unconventional discontinuous onset of the topological phases. 
}\label{fig:3}
\end{figure}


\section{The self-organized spin ladder}

Most of our numerical calculations are obtained for $N=70$, but in order to verify that the results would be valid in the thermodynamic limit, we also performed some finite size scaling with systems up to $N=200$. The preferred magnetic texture of this finite-size ladder is determined by minimization of the energy with respect to the allowed configurations of the localized moments ${\bm S}_{i,j}$.

Assuming the helical parametrization, we sweep through discretized values of $q$ and $\Delta q$ ($q_i=i\pi/100,\:\Delta q_j=j\pi/100,\:i,j=0,\ldots,100)$. For each pair $(q_i,\Delta q_j)$ the Hamiltonian (\ref{eq:hamil}) is numerically diagonalized, computing the ground state energy $E(q_i,\Delta q_j)$. In the next step, we search for the minimum of $E(q,\Delta q)$ for a range of model parameters, like $\mu, \Delta$, and $t_\perp$ determining $q_*$ and $\Delta q_*$. Using this algorithm we have computed the effective quasiparticle energies and the expectation values of physical quantities. Besides the perfect helical order along the legs, we have also checked the stability of dimerized configurations where $\phi_{i,j} = \phi_{i+1,j}$ for $i=1,3,5,\ldots$ and $j=1,2$. Stability of such block-spiral configurations has been suggested, e.g.~in Ref.~\cite{Herbrych16226}. To verify the validity of the helical ansatz we used the simulated annealing algorithm \cite{vanLaarhoven1987} based on Metropolis Monte Carlo approach to systems with mixed quantum-classical degrees of freedom \cite{PhysRevB.74.035109} [see appendix \ref{sec:sim_ann}].

Figs.~\ref{fig:phase}(a) and \ref{fig:phase}(b) show the stability diagrams of the system  with respect to $(\mu,\Delta)$. It can be noticed that the majority of stable configurations coincide with the ferromagnetic ($q_{*}=0$) or antiferromagnetic $q_{*}=\pi$ lateral orderings where $\Delta q_{*}=0$ or $\Delta q_{*}=\pi$, respectively. The regions of large chemical potential refer to the fully (or almost fully) filled bands with virtually no dependence on $(q,\Delta q)$ therefore these parts of the diagram do not develop well-defined magnetic textures. 

There exist, however, stable configurations with $q_*$ different from $0$ and $\pi$ located around the points marked by white crosses in Fig.~\ref{fig:phase}. In one of them $\Delta q_*=\pi$ and in the other one $\Delta q_*=0$. We have found there two types of helical ordering along the legs. These spirals are either identical on both legs or shifted by $\pi$, as schematically displayed in Fig.~\ref{scheme}. Furthermore, these regions reveal the stable topological phase hosting the zero-energy boundary modes. For an indirect proof we plot in Fig.~\ref{fig:phase}(c) the energy difference $\Delta E$ between two eigenstates right in the middle of the quasiparticle spectrum. To confirm that these zero energy states are MZMs we have also checked their Majorana polarization~\cite{Simon-2012,Sedlmayr-2015a,Sedlmayr-2015,Glodzik-2020} which can be probed by spin-polarized Andreev spectroscopy \cite{He-2014}, see Sec.~\ref{sec:polarization} for more details. Since the low-energy spectrum is symmetric, such a vanishing gap is a necessary condition for the zero-energy eigenstates. We observe that indeed in these regions $\Delta E=0$. What is interesting, within the topological region with $\Delta q_*=0$ there exists a small dome of stability of the dimerized phase, where the system is topologically trivial. It turns out that the two separate regions for $\Delta q_*=0$ and $\Delta q_*=\pi$ have quite different topological properties. We also discuss this issue from the symmetry point of view. Such two different topological regions stem from an additional degree of freedom in the ladder as compared to a single-leg chain. In the latter case, the only parameter characterizing the zero--temperature magnetic structure is the spiral pitch $q$. As regards the ladder, the relative mismatch between helical structures ($\Delta q_{*}$) turns out to be important as well.


\section{Topology and symmetry}

In order to facilitate our analysis of the Hamiltonian (\ref{eq:hamil}) we will first rewrite it using a gauge transformation, following which we can use a standard Fourier transform. We first take the step of writing the ladder degree of freedom using the Pauli matrices ${\bm\lambda}^i$. If ${\bm\sigma}^i$ and ${\bm\tau}^i$ stand for spin and particle-hole, then Hamiltonian \eqref{eq:hamil} becomes
\begin{align}
H_{s}=&
-t_{\parallel}\sum_i\Psi^\dagger_i{\bm\tau}^z\Psi_{i+1}+\rm{H.c.}
-t_{\perp}\sum_i\Psi^\dagger_i{\bm\tau}^z{\bm\lambda}^x\Psi_i
\nonumber\\&
-\mu\sum_i\Psi^\dagger_i{\bm\tau}^z\Psi_i
+\Delta\sum_i\Psi^\dagger_i{\bm\tau}^x\Psi_i
\\\nonumber&
+\frac{J}{4}\sum_i\Psi^\dagger_i\left({\bm S}_{i,1}\cdot\vec{\bm\sigma}[{\bm\lambda}^0+{\bm\lambda}^3]\right.
\\\nonumber&\qquad\qquad\qquad\left.
+{\bm S}_{i,2}\cdot\vec{\bm\sigma}[{\bm\lambda}^0-{\bm\lambda}^3]\right)\Psi_i \,,
\end{align}
with the convenient spinor notation
\begin{multline}
\Psi^\dagger_i= \\\left( c^\dagger_{i,1,\uparrow},c^\dagger_{i,2,\uparrow},c^\dagger_{i,1,\downarrow},c^\dagger_{i,2,\downarrow},c_{i,1,\downarrow},c_{i,2,\downarrow},-c_{i,1,\uparrow},-c_{i,2,\uparrow} \right)\, .
\end{multline}
In the following, we assume $\theta_i=\pi/2$ and $\phi_{i,j}=(i-1)q-(j-2)\Delta q$. The gauge transformation~\cite{Sedlmayr-2015} can be written explicitly as
\begin{equation}
    \mathbf{T}_{i,j}=\e^{-\im\frac{\phi_{i,j}}{2}\bsigma^z}\e^{-\im\frac{\pi}{4}\bsigma^y}\,,
\end{equation}
which gives $\mathbf{T}^\dagger_{i,j}\mathbf{S}_{i,j}\cdot\vec{\bsigma}\mathbf{T}_{i,j}=S\bsigma^z$. Introducing a Fourier transform $\mathcal{F}[H_s]=\sum_k\Psi^\dagger_k\mH_k\Psi_k$,  one then obtains
\begin{align}
\mH_k&=
f_{k,q}{\btau}^z+g_{\Delta q}{\btau}^z{\blambda}^x
-\frac{J}{4}{\bsigma}^z+\Delta{\btau}^x\\\nonumber&
+l_{k,q}{\btau}^z{\bsigma}^x+m_{\Delta q}{\btau}^z{\bsigma}^x{\blambda}^y\,.
\end{align}
The functions introduced are $f_{k,q}=-2t\cos k\cos q/2-\mu$, $g_{\Delta q}=-t\cos\Delta q/2$, $l_{k,q}=2t\sin k\sin q/2$, and $m_{\Delta q}=-t\sin\Delta q/2$.

\begin{figure}[t]
\begin{center}
\includegraphics[width=0.45\textwidth]{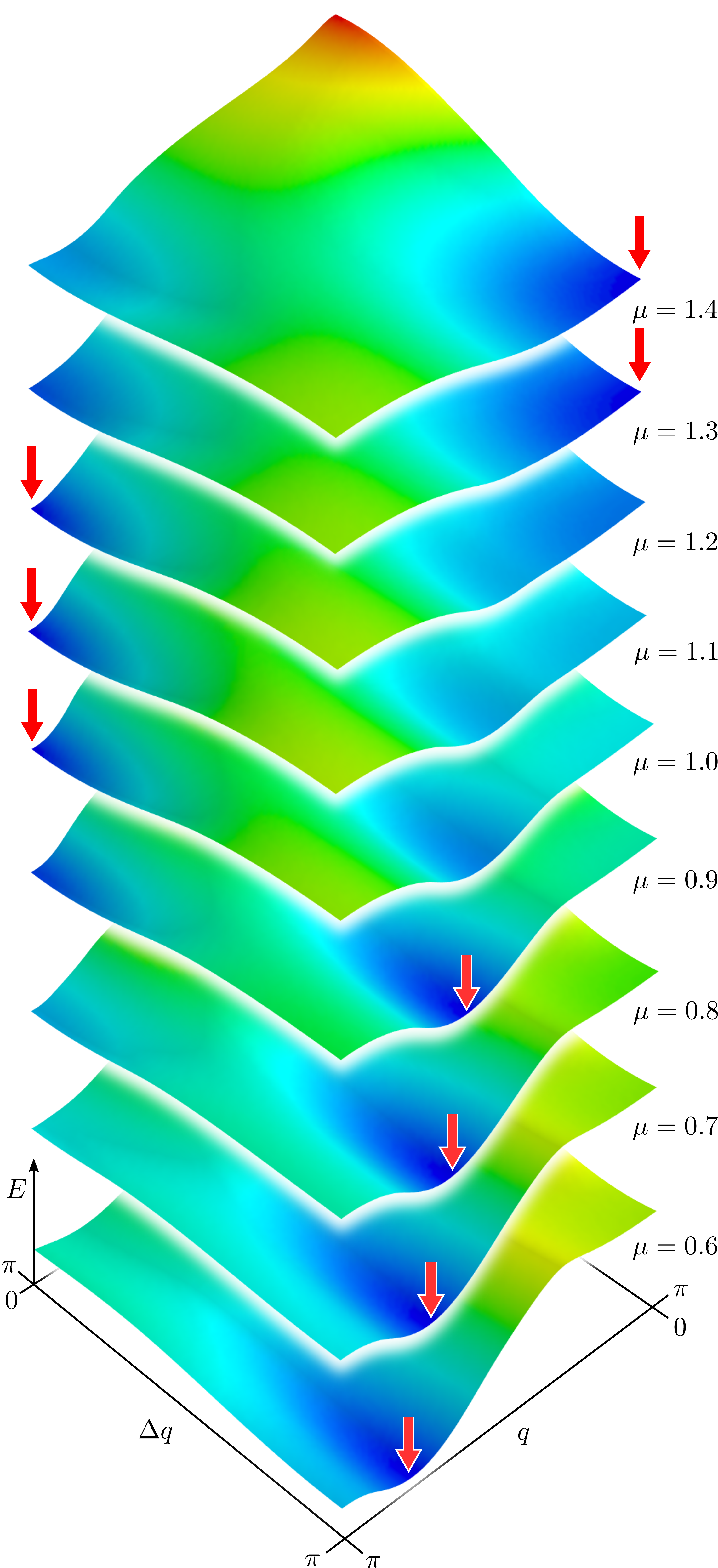}
\end{center}
\caption{Energy as a function of $q$ and $\Delta q$ for different values of the chemical potential and $\Delta=0.3$. The red arrows indicate $(q_*,\Delta q_*)$. The points in $(\mu,\Delta)$--phase diagram at which the energy landscapes were calculated are marked by black dots in Fig.~\ref{fig:phase}(a) and \ref{fig:phase}(b).}
\label{fig:landscape}
\end{figure}

With $\hat{K}$ being a complex conjugation, we potentially have the following symmetries:
\begin{itemize}
    \item[(i)] particle-hole symmetry $[\C,\mH]_+=0$, where $\C={\bm\tau}^y{\bm\sigma}^y\hat{K}$ and $\C^2=1$;
    \item[(ii)] time reversal symmetry $[\T_+,\mH]_-=0$, where $\T_+=\im{\bm\sigma}^z\hat{K}$ and $\T_+^2=1$;
    \item[(iii)] fermionic time reversal symmetry $[\T_-,\mH(\Delta q=\pi)]_-=0$, where $\T_-={\bm\sigma}^z{\bm\lambda}^y\hat{K}$ and $\T_-^2=-1$; and finally
    \item[(iv)] chiral symmetry: $[\V,\mH]_+=0$, where $\V=\C\T^+=-{\bm\tau}^y{\bm\sigma}^x$.
\end{itemize}
When $\T_-$ is present, there will be Kramer's pairs, but note that this symmetry acts in the ladder not the spin subspace. $\T_+$ can be destroyed by for example introducing non-planar spin densities. With both of these symmetries present, there is a further unitary symmetry $\mP=\T_+\T_-=-\im{\bm\lambda}^y$. We note here that neither of the time reversal symmetries used here correspond to the physical time reversal symmetry of electrons.

At this high symmetry point $\Delta q=\pi$ where this system has this additional unitary symmetry $[\mathcal{P},\mathcal{H}]=0$, we can first block diagonalize our Hamiltonian with respect to this symmetry operator:
\begin{equation}
    \tilde\mH_k=U^{-1}\mH_kU=
    \left(\begin{array}{cc}
         \mH_+&0\\0&\mH_-
    \end{array}\right)\,.
\end{equation}
In this basis $\tilde{\mP}=U^{-1}\mP U$ and
the rotation is given by
$U=\e^{-\im\frac{\pi}{4}\blambda^x}$.
 We note that as the gauge transformation $\mathbf{T}$ from the original basis to the local spin basis does not commute with this rotation $U$, in the original basis it would be given by
$U_o=\e^{\im\frac{\pi}{4}\blambda^y}$.
Following this rotation we find
$\tilde{\mH}_{\pm}=-\frac{J}{4}\bsigma^z+f_{k,q}\btau^z+(l_{k,q}\pm1)\btau^z\bsigma^x+\Delta\btau^x $.
It is straightforward to prove that this has only the chiral symmetry $\tilde{\V}=-\btau^y\bsigma^x$, and therefore is in class AIII with a $\mathbb{Z}$ invariant. A direct real space diagonalization shows that it has edge states, in complete agreement with the original Hamiltonian. Furthermore, one can use the chiral symmetry to calculate its topological invariant.

It therefore remains unclear if such MZM Kramer pairs could in principle be used for braiding operations. We note that within the strict classification scheme one should first block diagonalize the Hamiltonian with respect to all unitary symmetries. The combination of two time reversal symmetries leads to an additional unitary symmetry, and following block diagonalization, using the rotation $\rm{e}^{i\frac{\pi}{4}{\bm\lambda}^x}$, the resulting sub-block Hamiltonians have only chiral symmetry (class AIII) and therefore {\it a priori} there are no non-abelian MZMs.
Equivalently this can be seen in the full Hamiltonian by considering rotations within the degenerate Kramer pair subspace, only a specific set of bases refer to pairs of MZMs.

For now let us consider a general $\Delta q$, then we can find the $\mathbb{Z}$ invariant from~\cite{Gurarie2011}
\begin{equation}
    \nu=-\frac{1}{4\pi i}\int dk\, \textrm{tr}\,\mathcal{V}\partial_k\mH_k\left[\mH_k\right]^{-1}\,,
\end{equation}
where $\mathcal{V}=\C\T_+$ is the chiral symmetry operator. The integrand can be found analytically, but it is quite a long expression. As the chiral symmetry is still present even at the high symmetry point where $\mP=\T_+\T_-=-\im{\bm\lambda}^y$ is also a symmetry, this invariant also works in that case. This is an expression of the fact that one can consider such a system as BDI with an additional $T^{-}$ symmetry giving rise to Kramer's pairs \cite{Sedlmayr-2015JPCM}. We note that the extra TRS is not necessary for the invariant 2, it simply enforces a strict degeneracy on the MZM pairs.

The parity of the $\mathbb{Z}$ topological invariant~\cite{Sato-2009}, namely $\delta=(-1)^\nu$ can be found directly following Ref.~\cite{Sedlmayr-2016}. One obtains
\begin{equation}
 \delta=\text{sgn}(A_+A_-)
\end{equation}
where the explicit form for $A_\pm$ is given by
\begin{eqnarray}
A_\pm &=&  \frac{J^4}{2^8}-\frac{J^2}{8}\left[\cos (\Delta q)+\Delta ^2+\mu^2+2\right]
\nonumber \\
&+& \Delta ^4 + \mu ^4+6 \Delta ^2+2 \Delta^2 \mu^2+10 \mu ^2 
\nonumber \\ 
& \pm & 2 \mu  \left[4 \left(\Delta ^2+\mu ^2+2\right)-\frac{J^2}{4}\right] \cos \left(\frac{q}{2}\right) 
\nonumber \\ 
&+& \left(-\frac{J^2}{4}+4 \Delta ^2+12 \mu ^2+4\right) \cos q
\nonumber \\ 
&\pm &  8 \mu  \cos \left(\frac{3 q}{2}\right)
    +2 \cos 2 q+3 \, .
    \label{eq_12}
\end{eqnarray}
Its dependence on $\Delta$ and $\mu$ is presented in Fig. \ref{fig7}(a).


\section{Topological phase transitions}

An interesting fact about this ladder system is that it can have topological phase transitions which are accompanied by neither gap-closing nor symmetry changing. This can be traced back to the self-organized reordering of the spin structure, as the parameters are changed. Such discontinuous reorderings for $\Delta=0.3$ are clearly seen in Fig.~\ref{fig:3}. Additionally, evolution of the energy landscape in two of these transitions is illustrated in Fig.~\ref{fig:landscape}. By varying the chemical potential, the thick red arrow indicates discontinuous changes of the ground state configurations ($q_{*},\Delta q_{*}$) responsible for transitions to/from the topological phase.   

\begin{figure}[!htb]
\begin{center}
\includegraphics[width=0.48\textwidth]{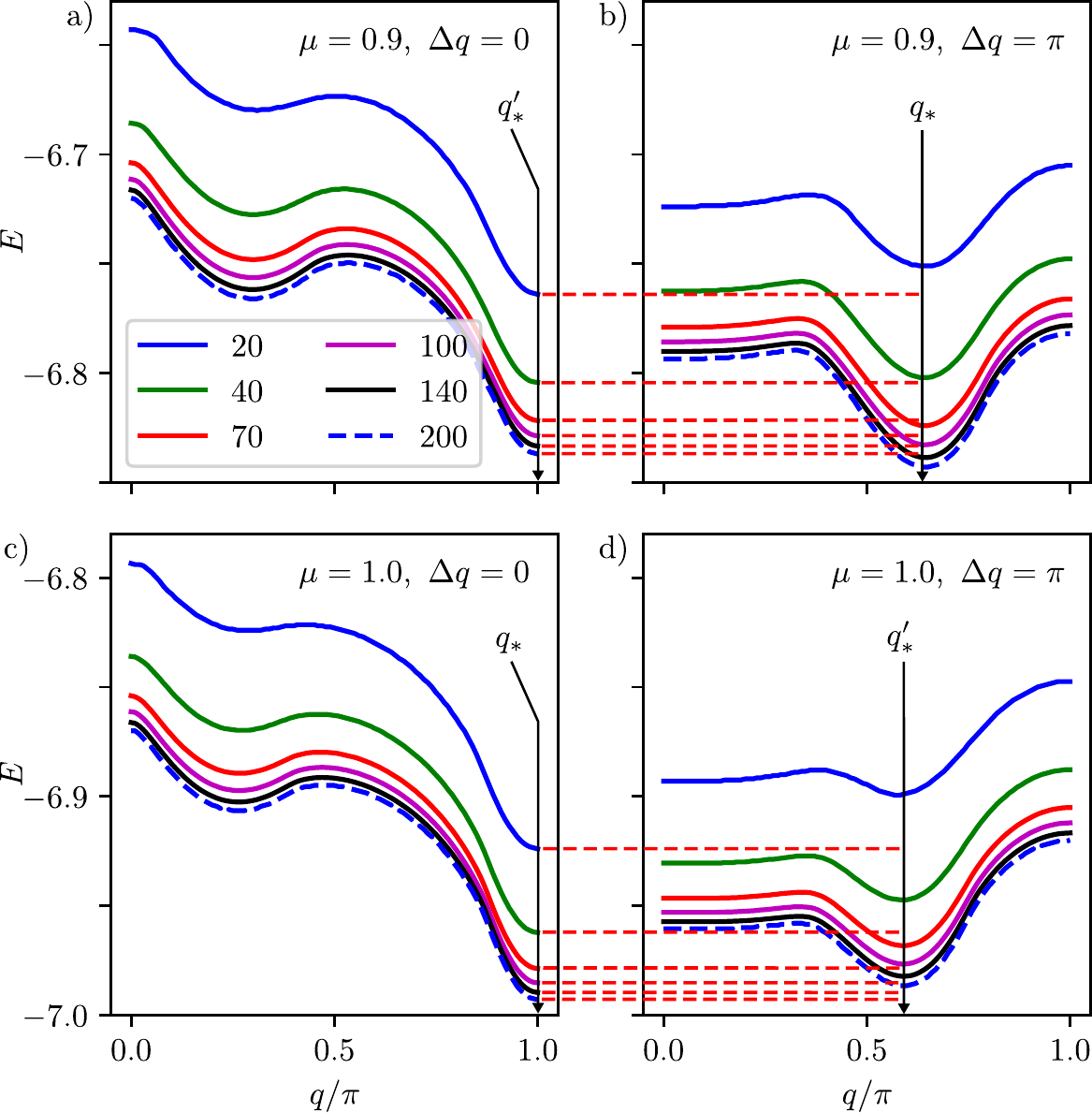}
\end{center}
\caption{Energy as a function $q$ for for $\Delta q=0$ (a and c) and $\Delta q=\pi$ (b and d) for $\mu=0.9$ (a and b) and $\mu=1.0$ (c and d) for different system sizes. Different lines correspond to $N=20,40,70,100,140$, and $200$, as indicated in the legend. The horizontal dashed red lines allows one to compare the depth of the global energy minimum at $q=q_*$ and the local minimum at $q=q'_*$.}
\label{fig:fss1}
\end{figure}

\begin{figure}[!htb]
\begin{center}
\includegraphics[width=0.48\textwidth]{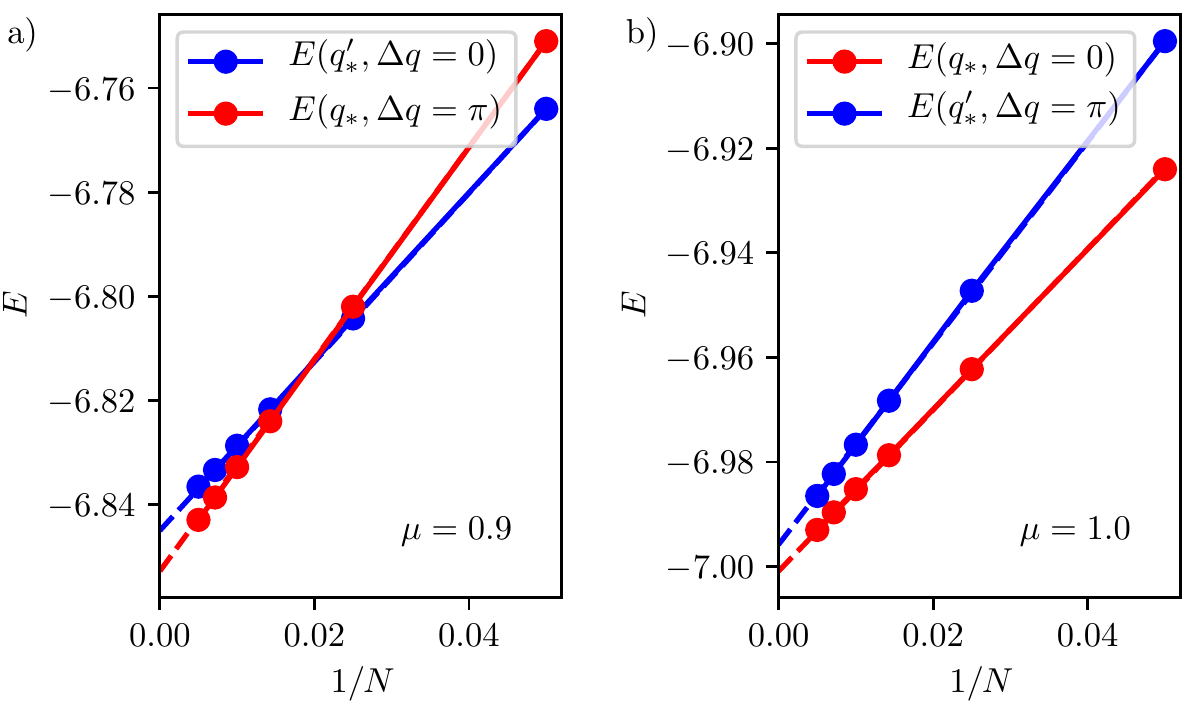}
\end{center}
\caption{Finite size scaling of energy at the global ($q_*$) and local ($q'_*$) energy minimum for $\mu=0.9$ (a) and $\mu=1.0$ (b). The points represent systems with lengths $N=20,40,70,100,140$, and $200$. The dashed lines show extrapolation $N\to\infty$.}
\label{fig:fss2}
\end{figure}

\begin{figure}
\begin{center}
\includegraphics[width=0.5\textwidth]{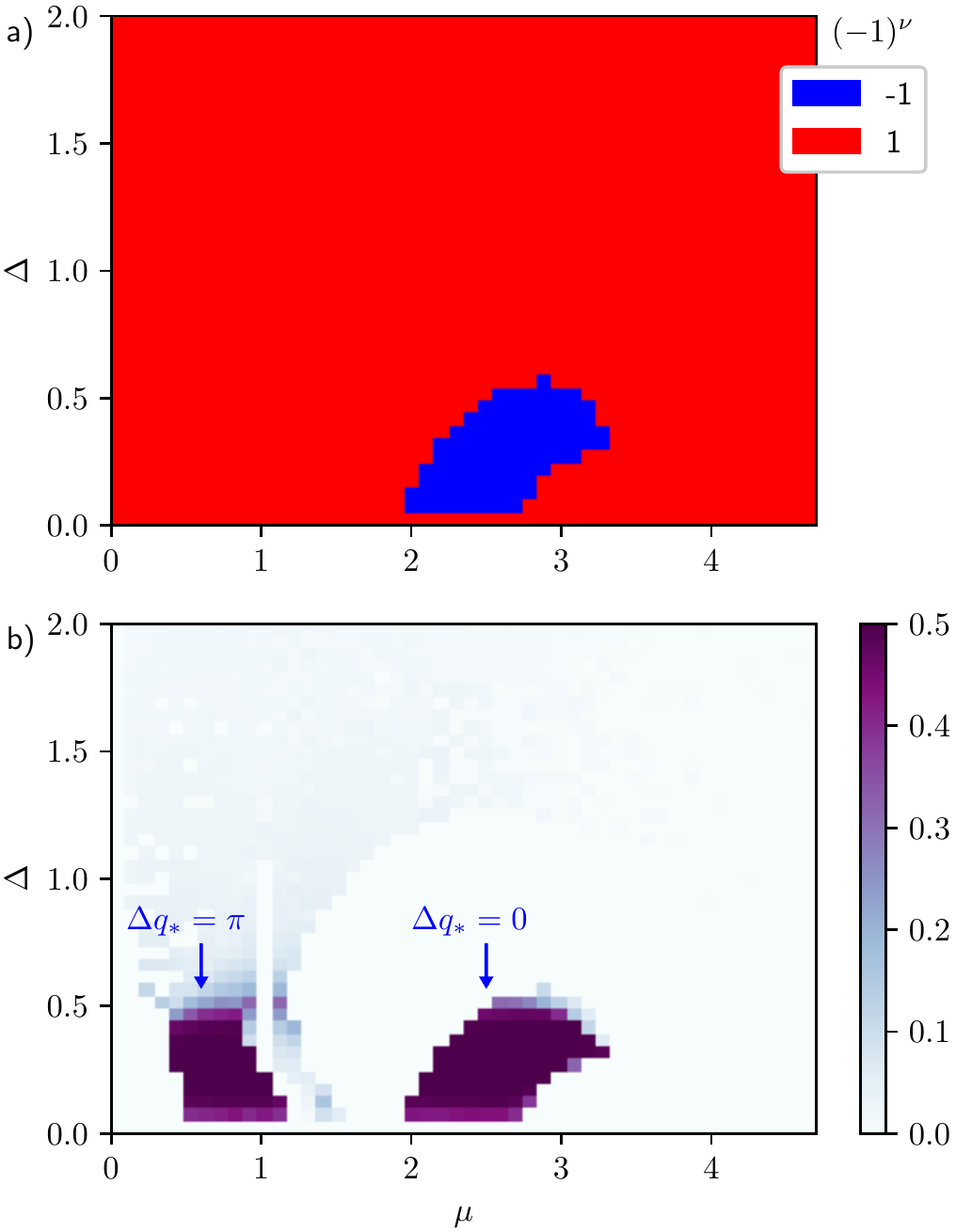}
\end{center}
\caption{The phase diagrams based on the parity calculation (a) and the Majorana polarization (b) obtained for system length $N=70$, both of which agree ideally with Fig.~\ref{fig:phase}.}\label{fig7}
\end{figure}

To demonstrate that this effect does not disappear in the thermodynamic limit we repeated the calculations for $\mu=0.9$ and $\mu=1.0$, i.e.~around one of the points where a discontinuous transition takes place, for systems with $N$ from 20 to 200 and performed a finite size scaling analysis. Fig.~\ref{fig:fss1} shows a comparison of the energy as a function of $q$ for $\Delta q=0$ and $\Delta q=\pi$ (left and right column, respectively) for $\mu=0.9$ and $\mu=1.0$ (upper and lower row, respectively). $q_*$ and $q'_*$ indicate the value of $q$ that corresponds to the global and local energy minimum respectively. One can notice there that for $N>40$ a discontinuous transition from $(q_*\approx 2,\Delta q_*=\pi)$ to $(q_*=\pi,\Delta q_*=0)$ occurs when the chemical potential increases from 0.9 to 1. Figure \ref{fig:fss2} shows that this holds true also in the thermodynamic limit $N\to\infty$. The absence of the transition for $N=20$ and $N=40$ indicates the importance of the finite size scaling.

\section{Majorana polarization}
\label{sec:polarization}

As an unambiguous check of the zero-energy quasiparticle modes emerging in the topological regions 
we also calculated, so called, {\it Majorana polarization} introduced in Ref.\ \cite{Simon-2012} 
as a suitable tool for probing the topological order parameter spread over region ${\cal{R}}$. Formally, it is defined by \cite{Sedlmayr-2015a,Sedlmayr-2015,Glodzik-2020}
\begin{equation}
    {\cal P}=\frac{1}{N}\sum_{i,j\in{\cal R}} \left( {\cal P}_{i,j\uparrow}-{\cal P}_{i,j\downarrow} \right) \,,
    \label{eq:polarization}
\end{equation}
where ${\cal P}_{i,j\sigma}=\langle\psi_\sigma|{\cal C}\hat{r}_{i,j}|\psi_\sigma\rangle $,
$\hat{r}_{i,j}$ is the projection onto site $i$ of $j$-th chain and ${\cal C}$ stands for the particle-hole operator. Since the zero-energy (Majorana) quasiparticles show up at the ends of the ladder, we choose ${\cal R}$ to be the half of all ladder sites $(i,j)$ for $i=1,\ldots,N/2$ and $j=1,2$. Fig.\ \ref{fig:phase}(d) confirms that the zero--energy states are indeed topological.

The magnitude $\left| {\cal P}_{i,j\uparrow}-{\cal P}_{i,j\downarrow} \right|$ of the local Majorana polarization can be probed by the spin-polarized Andreev spectroscopy \cite{He-2014}. It characterizes the particle-hole overlap of the emerging quasiparticles, whereas (\ref{eq:polarization}) probes their spacial coherence.

\section{Summary}

We have studied the self-organization of the classical magnetic moments of a finite-length ladder 
proximitized to a bulk superconductor. From numerical calculations we discover 
two different regions of possible topological phases (hosting the  zero-energy boundary modes) 
which are helically ordered along the legs with a characteristic pitch vector $q_{*}$ and 
a lateral mismatch $\Delta q_{*}=0$ or $\Delta q_{*}=\pi$. Analyzing the symmetry 
relations, we provided arguments for BDI or AIII classification of this system for $\Delta q_{*}=0$ or $\Delta q_{*}=\pi$ respectively, and discussed its  
$\mathbb{Z}$ topological invariant. We argue that the transition to the topological phase 
is unconventional, without any closing/reopening of the quasiparticle gap. We assign this unusual
behavior to a discontinuous changeover of the lateral shift $\Delta q_{*}$ between $0$ and $\pi$ 
upon varying the chemical potential. This effect should be observable experimentally with the use
of gate potentials. 
Similar discontinuous topological transitions might 
be possibly observed in other (non-superconducting) systems \cite{Juricic2017,Kore-2020}.


\begin{acknowledgments}
This research is supported by the National Science Centre (Poland) under the grants 2018/29/B/ST3/01892 (M.M.M.), 2019/35/B/ST3/03625 (N.S.), 2018/31/N/ST3/01746 (A.K.), and 2017/27/B/ST3/01911 (T.D.).
\end{acknowledgments}


\appendix

\section{Effect of perpendicular hopping}
\label{sec:perpendicular}

\begin{figure*}[t!]
\begin{center}
\includegraphics[width=\textwidth]{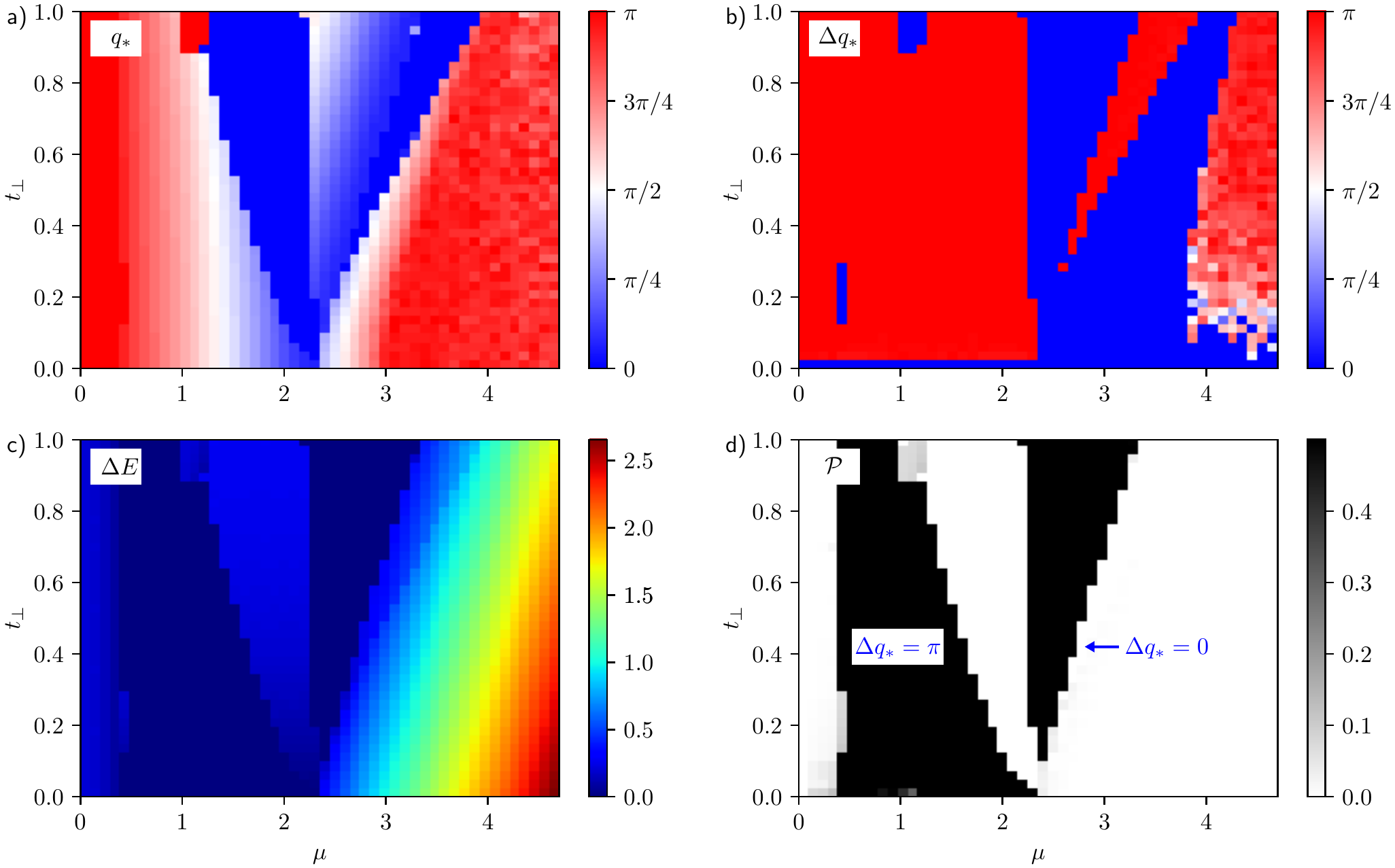}
\end{center}
\caption{The same as in Fig.\ \ref{fig:phase} with respect to the  perpendicular hopping integral $t_{\perp}$ obtained for $\Delta=0.3$.}
\label{fig:phase-perp}
\end{figure*}

For additional insight into the topological superconducting phase of the self-organized magnetic ladder we also examined the stable magnetic structures allowing the inter-leg coupling, $t\to t_{\perp}$, to vary. Fig.~\ref{fig:phase-perp} displays the results obtained for $\Delta=0.3$. The inter-leg hopping, $t_{\perp}$, varies from $0$ (decoupled chains) to $1$ (identical hopping along and across the legs). Let us remark that for $t_{\perp}=0$ the pitch vector $q_*$ and the chemical potential $\mu$ corresponding to the topologically nontrivial ground state perfectly agrees with the previous results obtained for a single chain \cite{Maska-2019}, whereas $\Delta q_*$ is then  meaningless. 
Upon increasing the coupling $t_{\perp}$  the initial topological phase (appearing at $\Delta q_{*}=\pi$ for $t_{\perp}=0$) gradually shrinks and an additional topological phase is established (at $\Delta q_{*}=0$), starting from $t_{\perp}\approx 0.1$.


\section{Simulated annealing}
\label{sec:sim_ann}

\begin{figure}[htb]
\begin{center}
\includegraphics[width=0.4\textwidth]{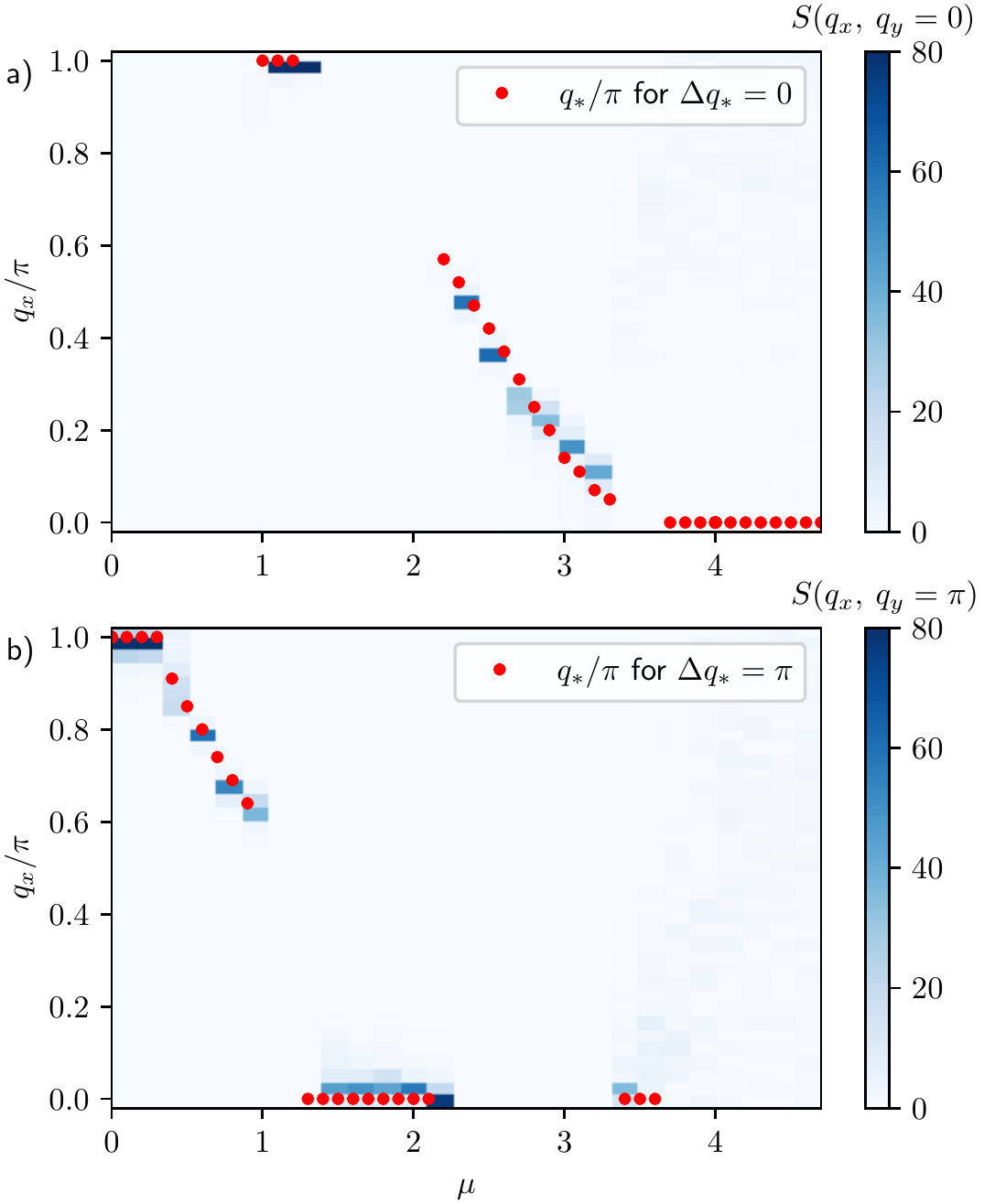}
\end{center}
\caption{Comparison of the results obtained under the assumption of a regular helical magnetic structure and unrestricted results obtained with the simulated annealing algorithm. The blue color in panel (a) shows a false color map of the static spin structure factor $S(q_x,q_y=0)$ and in panel (b) $S(q_x,q_y=\pi)$. The red points show $q_*$ and $\Delta q_*$. For $\mu\gtrsim 3.6$ the ordering temperature is too low to be reached in the SA which explains the disagreement in this region. It can also be seen in Fig.~\ref{fig:phase}, where the finite numerical accuracy does not allow for an unambiguous determination of the ground state spin configuration.}
\label{fig:sim_ann}
\end{figure}

Most of the presented calculations have been obtained under the assumption that the magnetic structure can be parameterized as ${\bm S}_{i,j} = S\left(\cos\phi_{i,j},\:\sin\phi_{i,j},\:0 \right)$ where $\phi_{i,1}  = iq$, $\phi_{i,2} = iq + \Delta q$. To verify the validity of this ansatz, we used the simulated annealing (SA) algorithm to generate the actual zero temperature spin configurations which correspond to the global minimum of the ground state energy \cite{vanLaarhoven1987}. In this approach we start with high temperature and random spin configurations. Then, we perform Metropolis Monte Carlo (MC) simulations during which the temperature is gradually reduced. In each MC step a new spin configuration is tried, the Hamiltonian \eqref{eq:hamil} is diagonalized, and then the new configuration is accepted or rejected on the basis of the change of the free energy \cite{PhysRevB.74.035109}.
To avoid trapping in a local energy minimum, the temperature decreases in a sawtooth-like pattern, where at the end of each step the system is reheated before further linear temperature reduction. When the spin configurations are close to the global minimum of the ground state energy, we calculate the static spin structure factor
\begin{equation}
    S({\bm q})=\dfrac{1}{N^2}\sum_{k,l}\langle {\bm S}_k\cdot {\bm S}_l\rangle e^{i{\bm q}\cdot {\bm r}_{kl}}\,,
    \label{eq:sq}
\end{equation}
where $k,l$ denote the spin coordinates along the $x$ and $y$ directions, i.e., $k=(i,j),\:l=(i',j')$ and ${\bm r}_{kl}$ is the position vector from site $k$ to $l$. $\langle\ldots\rangle$ represents the thermal average over the ensemble generated in the MC runs. As $T\to 0$ fluctuations vanish and $\langle {\bm S}_k\cdot {\bm S}_l\rangle \to \left.{\bm S}_k\cdot {\bm S}_l\right|_\mathrm{min}$, where ``min'' denotes the configuration that minimizes the ground state energy. For a perfect spiral ordering, $S({\bm q})$ has a peak at $q_x=q_*$ and $q_y=\Delta q_*$ with its magnitude dependent on the spiral pitch.

\begin{figure}[ht!]
\begin{center}
\includegraphics[width=0.4\textwidth]{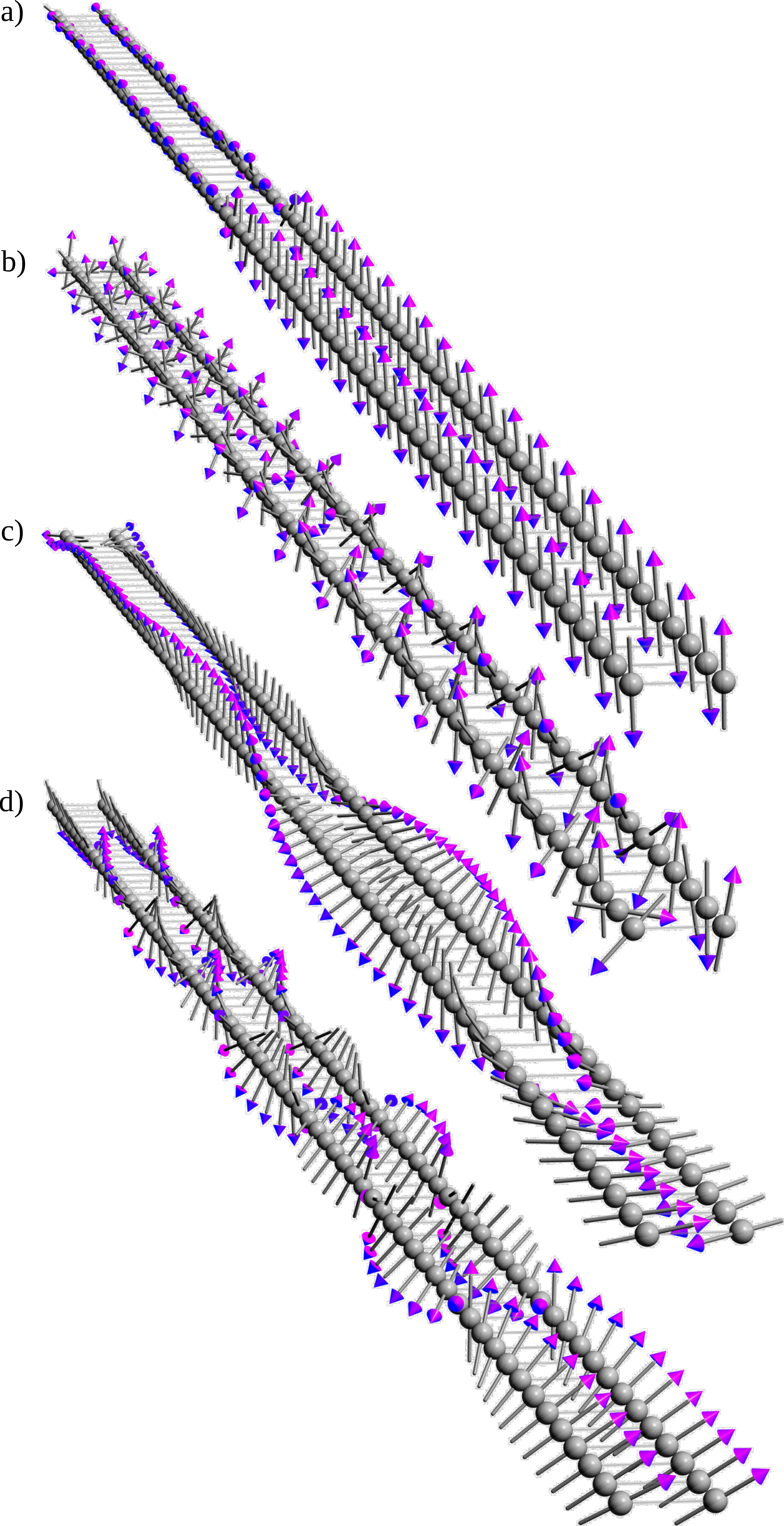}
\end{center}
\caption{Examples of spin configurations obtained within the SA approach for $\mu=0.2$ (a), $0.6$ (b), $1.8$ (c), and $3.2$ (d). In all cases $\Delta=0.3$ is assumed.}
\label{fig:sim_ann_examples}
\end{figure}

Fig.~\ref{fig:sim_ann} shows a comparison of $S({\bm q})$ for an unrestricted magnetic structure obtained in the SA and $q_*,\Delta q_*$. 
Since the size of the system in the $y$ direction is only 2, there are only two allowed values of $q_y=0$ and $q_y=\pi$. These two slices of $S({\bm q})$ are presented in Fig.~\ref{fig:sim_ann} in panels (a) and (b), respectively. Some slight discrepancies result from imperfections in the configurations generated in SA. Examples of the configurations are presented in Fig.~\ref{fig:sim_ann_examples}.

\begin{figure}[ht]
\begin{center}
\includegraphics[width=0.5\textwidth]{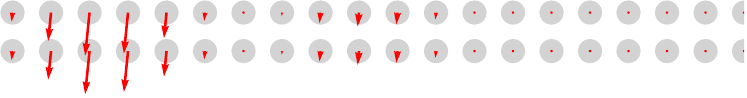}
\end{center}
\caption{Local polarization ${\cal P}_{i,j\uparrow}-{\cal P}_{i,j\downarrow}$ for $\mu=3.2$ and $\Delta=0.3$. Only half of the ladder is shown.}
\label{fig:polar_3.2}
\end{figure}

Panel (a) shows a single point defect (domain wall) that eventually would vanish after longer SA. However, since the SA was aimed only at verifying the assumed ansatz, the calculations were not performed for a very long time. Most of the configurations are, however, very close to perfect spirals, as demonstrated in panel (b). Small long-wavelength fluctuations sometimes are superimposed on a regular order, especially close to discontinuous transitions, as shown in panel (c). In a narrow range of the model parameters, the spin configurations are affected by the presence of the edges of the ladder. As can be seen in panel (d), close to the ends of the ladder the spiral order is replaced by a ferromagnetic alignment of the spins. Since the spiral is necessary for the topological phase, this effect can be detrimental to Majorana modes, which are supposed to be located in this very region. To examine if this is actually the case, we have calculated the spatial distribution of the local Majorana polarization ${\cal P}_{i,j\uparrow}-{\cal P}_{i,j\downarrow}$. The results presented in Fig.~\ref{fig:polar_3.2} show that the Majorana modes are still located close to the ends of the ladder, despite the fact that the spiral order is strongly suppressed in these regions.

\section{Quasiparticle spectra}
\label{sec:eignestates}

\begin{figure}
\includegraphics[width=0.425\textwidth]{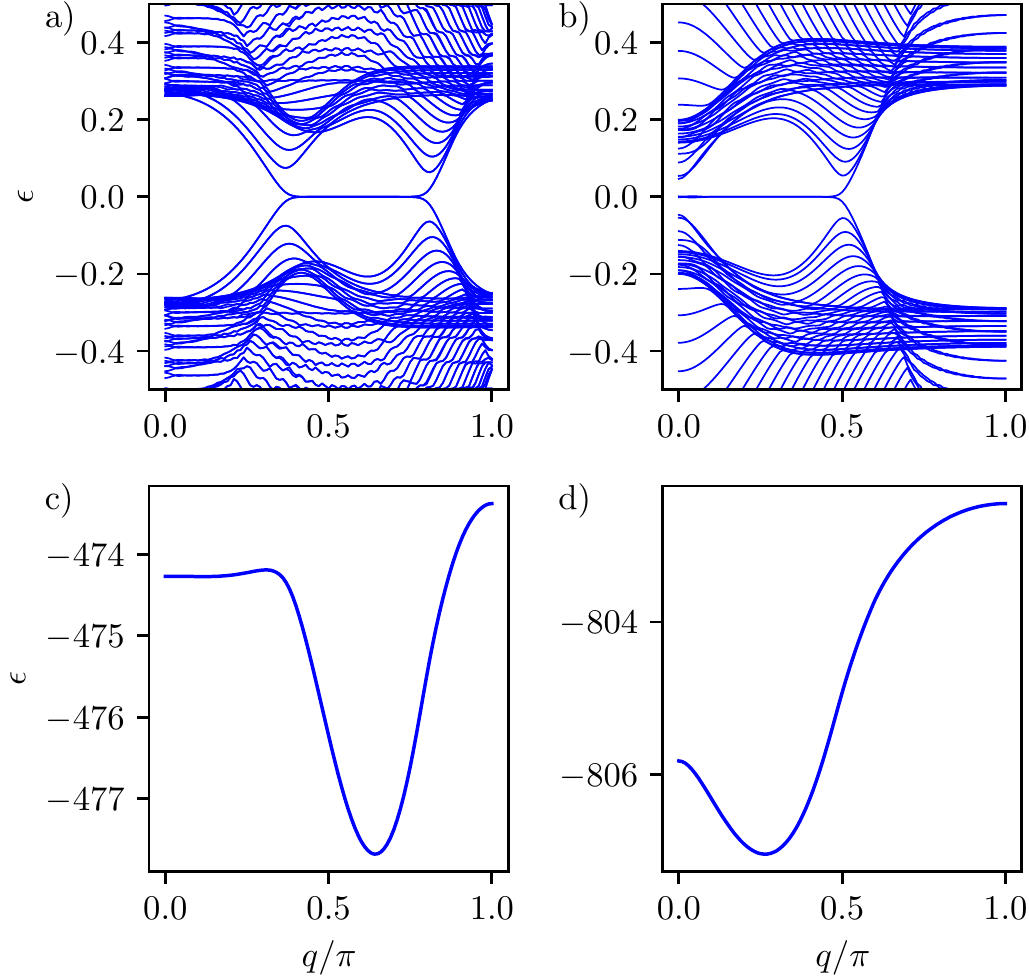}
\caption{The upper row: The quasiparticle energies for $\mu=0.9$, $\Delta  q=\pi$ (a) and for $\mu=2.8$, $\Delta q=0$ (b). In both cases  $\Delta=0.3$ is assumed.
The bottom row: The corresponding ground state energy as a function of $q$ for $\mu=0.9$, $\Delta  q=\pi$ (c) and $\mu=2.8$, $\Delta q=0$ (d).}
\label{fig:spectr1}
\end{figure}

In this appendix we present the typical quasiparticle energies 
as functions of the pitch vector $q$ obtained numerically for $\Delta=0.3$ and two 
values of the chemical potential, $\mu=0.9$ (Fig.~\ref{fig:spectr1}(a,b)) and $\mu=2.8$ 
(Fig.~\ref{fig:spectr1}(c,d)). These values of $\mu$ correspond to the points in the topologically non-trivial phases (marked by the white crosses in Fig.~\ref{fig:phase}) 
predicted in the main part of our paper. In the bottom panels in both figures we show the total energy of our system. One can clearly see that indeed the ground state configurations (at pitch vector $q_{*}$) 
coincide with the topologically non-trivial region hosting  the zero-energy modes. Such a tendency 
for the topological ground state resembles {\it topofilia}, predicted previously for 
self-organized magnetic chains~\cite{Vazifeh-2013,Simon-2013,Braunecker-2015,Klinovaja-2013,Paaske-2016,Scalettar-2017,Maska-2019}.

%


\end{document}